# First-order spatial coherence measurements in a thermalized two-dimensional photonic quantum gas

Tobias Damm[1], David Dung[1], Frank Vewinger[1], Martin Weitz[1] & Julian Schmitt[1]

[1]*Institut für Angewandte Physik, Universität Bonn, Wegelerstr. 8, 53115 Bonn, Germany*

**Phase transitions between different states of matter can profoundly modify the order in physical systems, with the emergence of ferromagnetic or topological order constituting important examples. Correlations allow to quantify the degree of order and classify different phases. Here we report measurements of first-order spatial correlations in a harmonically trapped two-dimensional photon gas below, at, and above the critical particle number for Bose-Einstein condensation, using interferometric measurements of the emission of a dye-filled optical microcavity. For the uncondensed gas, the transverse coherence decays on a length scale determined by the thermal de Broglie wavelength of the photons, which shows the expected scaling with temperature. At the onset of Bose-Einstein condensation true long-range order emerges, and we observe quantum statistical effects as the thermal wave packets overlap. The excellent agreement with equilibrium Bose gas theory prompts microcavity photons as promising candidates for studies of critical scaling and universality in optical quantum gases.**

## Introduction

Different states of matter are universally characterized by the type of order, which is encoded in their correlation properties[1]. A gas of massive particles at high temperatures or low densities, for example, exhibits short-range spatial first-order correlations as inherited by its classical single-particle properties[2–4]. The latter are determined by the thermal de Broglie wave describing a particle of (effective) mass $m$, which is defined as $\lambda_{\text{th}}=h/\sqrt{2\pi m k_B T}$, where $T$ denotes the temperature, equaling the de Broglie wavelength of a particle at an average thermal velocity[5]. Quantum statistical effects emerge in a cold and dense gas when the thermal de Broglie wavelength exceeds the mean interparticle spacing. For massive bosonic particles, as atoms with integer spin, as well as for two-dimensional gases of photons or polaritons, Bose-Einstein condensation (BEC) into a macroscopically occupied ground state has been observed[6–13]. Upon reaching Bose-Einstein condensation, the achieved macroscopic ground state population enhances the coherence length far beyond the thermal de Broglie wavelength, essentially covering the complete sample size[3,5]. This has been strikingly confirmed both in the thermal and condensed phase regime for massive particles, as ultracold atomic gases[14–19].

For optical quantum gases as photons and exciton-polaritons an effective mass is established by tailoring the dispersion relation of photons within optical microcavities[20]. Here, upon condensation long-range order has also been revealed, similarly indicating the spontaneous onset of a new phase[9–12,21–24]. For such two-dimensional systems at high phase-space densities, in principle both BEC and Berezinskii-Kosterlitz-Thouless (BKT) type of phases may occur, where the former is associated with true long-range order and the latter exhibits first-order correlations decaying algebraically in space[25–34]. In previous exciton-polariton experiments, however, the observed transverse coherence below and partly also above the condensation threshold was limited by the finite cavity linewidth[21,31]. More recently, Marelic et al. reported on evidence for spatial coherence in a harmonically trapped photon gas, without finding quantitative agreement with Bose gas theory for all particle numbers, as attributed to the finite spatial resolution of the used imaging system and the nonequilibrium nature of the studied optical system far above the condensation threshold[24]. Indeed, the genuine thermal de Broglie wavelength of such two-dimensional photonic gases so far has remained elusive.

Here we report on a quantitative study of the first-order spatial correlations of a two-dimensional harmonically trapped photon gas at equilibrium conditions, both in the classical and in the Bose-Einstein condensed phase using optical interferometry. For this, we have developed an experimental platform excelling in high sensitivity and spatial resolution of more than an order of magnitude below the width of the condensate mode. In the uncondensed phase, we directly observe the thermal de Broglie wavelength through the correlation length of the photons, which has both the expected absolute value and the expected temperature scaling. As the photon number is increased, long-range order, as indicated by a significant increase of the transverse coherence length, emerges when the mean distance between the optical wave packets approaches the measured thermal de Broglie wavelength. Our direct look at the coherence properties of the photons from the interferometrically measured in-plane first-order correlations verifies the good applicability of thermodynamic ideal Bose gas theory for the present photonic quantum gas.

## Results

**Preparation and Characterizing Measurements**

To prepare a two-dimensional photon gas we use a microcavity setup filled with a liquid dye solution (see the left-hand side of Fig.1a). Here, the photons are confined by two highly reflective curved mirrors spaced by a distance in the wavelength regime. The correspondingly large free spectral range in the microcavity restricts the dye molecules to in good approximation emit only into transversal cavity modes belonging to one fixed longitudinal mode number. The photon gas then becomes two-dimensional with the lowest photon energy at $\hbar\omega_{cutoff} \simeq 2.1$eV for the transversal ground mode, introducing an effective mass of the cavity photons in our experiment. The resulting quadratic optical dispersion, see eq. (1) below, is the same as for a massive particle[35]. Figure 1b shows the measured energy-momentum-relation for the uncondensed two-dimensional photon gas (see Methods). The data follow a quadratic scaling with the transverse wave vector $k_r$ with the dashed line showing the expected dispersion for an effective photon mass $m_{eff} = \hbar\omega_{cutoff}/c^2 \simeq 7.76(2) \times 10^{-36}$kg, where $c=c_0/n$ denotes the speed of light in the medium with refractive index $n$, as derived from the cavity parameters. The agreement with the experimental data verifies the predictions for a quadratic (non-relativistic) dispersion in the weak coupling regime[36,37]. Given the non-vanishing effective photon mass, the concept of a thermal de Broglie wavelength can be extended to the cavity photons. Spatially, the mirror curvature induces a harmonic trapping potential of trapping frequency $\Omega$ for the photon gas, yielding a photon energy in the cavity

$$E \simeq m_{eff}c^2 + \frac{(\hbar k_r)^2}{2m_{eff}} + \frac{1}{2}m_{eff}\Omega^2 r^2. \qquad (1)$$

To achieve thermalization, the photons are coupled to a dye solution in the microcavity (Rhodamine 6G solved in ethylene glycol) by repeated absorption re-emission cycles.

The two transverse modal degrees of freedom thermalize to the (internal rovibrational) temperature $T$ of the dye, leading to Bose-Einstein distributed photon energies of order $\sim k_\mathrm{B} T$ above the low-energy cutoff provided that the thermalization is sufficiently fast[35,38]. To generate a photon gas of total particle number $N$ and compensate for optical losses from e.g. mirror transmission, the dye microcavity is pumped by a laser beam (see Methods). In the presence of the harmonic trapping potential, the two-dimensional photon gas exhibits a phase transition to a Bose-Einstein condensate at non-vanishing temperatures or finite particle numbers, correspondingly. Above a critical particle number of $N_\mathrm{c}=\pi^2/3\,(k_\mathrm{B}T/(\hbar\Omega))^2 \approx 94000$ photons at room temperature ($T$=300K), with the harmonic trapping frequency $\Omega/(2\pi) \approx 37$GHz, Bose-Einstein condensation has been observed[12,13,39], and the temporal evolution into equilibrium has been studied[40,41].

**Interferometric Setup and Model**

To investigate the coherence properties of the photon gas we employ the interferometric measurement schematically depicted in Fig.1a, see also Methods for details. The emission from the dye microcavity is collimated and after passing a polarizer to lift the polarization degeneracy sent to a Michelson interferometer with a movable cat-eye retroreflector replacing one of the mirrors[9,31]. The plane mirror in the reference arm is slightly tilted leading to a fringe-type interference pattern, which we read out with a camera, see Fig.1c for an example. As the retroreflector inverts the image, each point $\mathbf{r}=(x,y)$ in the camera plane corresponds to the interference of fields of the cavity emission at two points at transverse positions $\mathbf{r}$ (reference path) and $-\mathbf{r}$ (cat-eye path) prior to entering the interferometer. At the detector, we expect an interference pattern of the form

$$I_\mathrm{d}(\mathbf{r}) = \tfrac{1}{4}\left[I(\mathbf{r}) + I(-\mathbf{r}) + 2\sqrt{I(\mathbf{r})I(-\mathbf{r})}\cos(\Delta\phi)\left|g^{(1)}(\mathbf{r},-\mathbf{r};\tau)\right|\right], \quad (2)$$

where $I(\mathbf{r})$ corresponds to the intensity distribution of the cavity emission at transverse position $\mathbf{r}$. Further, $\Delta\phi=\arg[g^{(1)}(\mathbf{r},-\mathbf{r};\tau)]$ and $\tau=\Delta\ell/c_0$ denotes the time delay accumulated between arms due to a path length difference $\Delta\ell$. The normalized first-order spatial correlation is defined as[42]

$$g^{(1)}(\mathbf{r},-\mathbf{r};\tau) = \frac{\langle \mathbf{E}^+(\mathbf{r},t)\mathbf{E}(-\mathbf{r},t+\tau)\rangle}{\sqrt{\langle|\mathbf{E}(\mathbf{r},t)|^2\rangle}\sqrt{\langle|\mathbf{E}(-\mathbf{r},t+\tau)|^2\rangle}}, \quad (3)$$

where $\mathbf{E}(\mathbf{r},t)$ denotes the quantized electric field operator at transverse position $\mathbf{r}$ and time $t$ and brackets account for (ensemble) averaging under stationary statistics (see

Supplementary notes 1-4). Experimentally, we use the fringe visibility $V=(I_{max}-I_{min})/(I_{max}+I_{min})$ as a measure for the correlation function, where $I_{max}$ ($I_{min}$) denote maximum (minimum) intensities of the fringe pattern around the path difference $\Delta\ell$. The visibility is related to the first-order coherence following

$$V(\mathbf{r},\tau) = \frac{2\sqrt{I(\mathbf{r})I(-\mathbf{r})}}{I(\mathbf{r})+I(-\mathbf{r})} \left|g^{(1)}(\mathbf{r},-\mathbf{r};\tau)\right|. \qquad (4)$$

Correspondingly, when varying $\tau$ by changing the longitudinal position of the cat-eye retroreflector with a piezo-driven stepper motor translator, we can extract both longitudinal and transverse coherence properties of the emission from the dye microcavity. Intuitively, the interference signal at each point $\mathbf{r}$ corresponds to the interference expected from a Young's double slit experiment[21,22] with a slit separation of $2|\mathbf{r}|$, as seen when inspecting the optical paths in Fig.1a. Figures 1(c,f) show spatial interference fringes for a fixed path length difference (near $\Delta\ell=0$) for the case of a photon Bose-Einstein condensate, while Figs.1(d,e,g) show images along with a cut along their $x$-axes at $y=0$ recorded far below the threshold for condensation. For the latter, two path length differences have been selected that correspond to a phase shift of $\pi$. The images indicate the large difference in the transverse coherence length for the case above and below threshold.

We find the expected correlation function eq. (3) for the case of the two-dimensional photon gas in a harmonic trap similar to earlier work[4,43]. Briefly, to find $g^{(1)}(\mathbf{r},-\mathbf{r};\tau)$, we expand the electric field operators in eigenfunctions of the harmonic oscillator and assume that the photon gas is in thermal equilibrium at temperature $T$. The resulting equations can then be solved numerically, where we use the eigenfunctions of the 700 lowest harmonic oscillator modes; details are summarized in the Supplementary information. Far below threshold, the distribution function is well described by a Boltzmann distribution, which allows for an analytic solution of the spatial correlation function at equal times, $g^{(1)}(\mathbf{r},-\mathbf{r};0)=\exp(-4\pi|\mathbf{r}|^2/\lambda_{th}^2)$. Correspondingly, a measurement of the spatial correlation length, defined as $\ell_c = \lambda_{th}\sqrt{\ln 2/(4\pi)}$ (FWHM), allows to directly determine the thermal de Broglie wavelength. In our numerical calculations, we account for the noise floor of the used camera, which reduces the measured interference contrast especially in regions with low intensities.

**Temporal and Transverse Coherence**

First, we have studied the temporal coherence by scanning the longitudinal position of the cat-eye retroreflector and studying the interference pattern near **r**=0. A corresponding temporal fringe pattern is displayed in Fig.2a (top), showing the signal detected by one camera pixel versus the temporal delay due to path length difference of the Michelson interferometer. The bottom panel shows the corresponding variation of the fringe visibility. The visibility reduces for large path length differences, and from the decay we find a coherence time in the uncondensed phase of $\tau_c \simeq 129(6)$fs at a temperature of $T$=297K. This value exceeds the expected coherence time of $\sqrt{3}\hbar/(k_B T) \simeq 44$fs, which we attribute to the finite imaging resolution of our setup (see Supplementary note 5). Indeed, when including an averaging over an area given by the point spread function in the numerical calculations, the results agree much better with the measured data, demonstrating qualitatively the strong influence of finite spatial resolution on the measured temporal coherence data. In accordance with theory predictions, the first-order correlation time of the uncondensed photon gas is more than four orders of magnitude shorter than values observed in the Bose-Einstein condensed phase with heterodyne measurements[23], and slightly smaller than other reported results[24].

To study the transverse coherence, Fig.2b gives the observed visibilities of the central fringes ($\Delta\tau \simeq 0$ in Fig.2a) versus transverse position, directly providing a spatial map of the coherence of the thermal photon gas. To extract the coherence length, we subtract a visibility offset given by the noise characteristics of the camera (see Supplementary note 6) and radially average the visibility. Fig.2c gives the corresponding variation versus transverse distance from the origin. The visible rapid decay with increasing distance is understood in terms of the limited transverse coherence of the thermal photon gas. After correcting for the spatial resolution of the imaging system, measured using a SNOM-fiber of <200nm aperture diameter placed in the cavity plane (see Methods), we can readily determine the spatial correlation length of the two-dimensional, harmonically confined gas. From this we extract the corresponding thermal de Broglie wavelength $\lambda_{th}$=1.48(1)μm. This agrees well with theoretical value of 1.482(2)μm, obtained by using the above quoted effective photon mass and its uncertainty in the formula for the thermal de Broglie wavelength. In contrast to a determination of the de Broglie wavelength by momentum-resolved emission spectra[31], we here directly observe the spatial coherence of the photon wave packet.

To test for the expected temperature scaling of the thermal de Broglie wavelength, the cavity was heated using two electric heaters placed on the side of the cavity mirrors. This allows us to tune the temperature by some 70K, as at higher temperatures the solvent starts to noticeably evaporate. Figure 3 shows the variation of the experimentally determined thermal de Broglie wavelength with temperature of the dye microcavity. With increasing temperature, the observed transverse coherence length and correspondingly the extracted de Broglie wavelength shortens. The shown error bars give the size of the statistical uncertainty and the grey line the systematic uncertainty due to correction for the point spread function of the imaging system. A fit to the data yields a variation $T^{-0.51(3)}$, which is in very good agreement with the predicted $1/\sqrt{T}$ scaling.

In a next step, the variation of the transverse coherence of the photon gas with increasing photon number was studied. At the onset of Bose-Einstein condensation, we expect a sharp increase in coherence length, as is qualitatively already visible from inspecting Figs.1(c,f). To quantify this, we measured spatial maps of the fringe visibility for different ratios of the total photon number $N$ and critical photon number $N_c \simeq 94000$, see Fig.4a. The corresponding variation of the fringe visibility with radial distance from the center is given in Fig.4b, demonstrating good agreement with numerical calculations for corresponding values of $N/N_c$ accounting for detection noise floor and uncertainties in the measured photon number (shaded areas). In this parameter regime, continuous operation of the photon gas is not feasible due to excitation of long-lived triplet states of the dye molecules. We therefore use pump pulses of 600ns temporal length, which is more than two orders of magnitude above the thermalization timescale[40]. This pulsed operation has not been observed to affect the degree of thermalization, but it leads to a signal-to-noise ratio below that of the measurements of the uncondensed gas (Fig.2). From our data, we find the onset of condensation at a phase-space density $\tilde{n}_c \lambda_{th}^2 = 3.2(6)$ with the critical central density $\tilde{n}_c$ (see Methods), see Fig.4c, which corresponds within the uncertainties to the expected value $\tilde{n}_c \lambda_{th}^2 = \pi^2/3$ at criticality[27,44]. In the condensed phase, the coherence length strongly increases and soon exceeds the condensate diameter (FWHM, horizontal solid line in Fig.4c). For values $N/N_c \ll 1$, corresponding to the thermal regime, the coherence length as expected approaches $\lambda_{th}\sqrt{\ln 2 /(4\pi)}$. When comparing our data to the numerically derived coherence function, we see a good agreement both below and above the condensation threshold when we take the camera characteristics into account (solid line in Fig.4c).

Finally, we in more detail investigate the photon gas correlations in a regime close to the condensation threshold where quantum degeneracy is reached. Figure 5 gives the three measured data sets with lowest photon numbers (from Fig.4b) in the condensed phase regime, fitted with numerically calculated visibility curves using $N/N_c$ as a fitting parameter. Our previous discussion has identified two distinct regimes: firstly, we find correlations in the uncondensed photon gas in good agreement with a Gaussian decay on a length scale given by the thermal de Broglie wavelength (Fig.2), and secondly, coherence exceeds the ground mode diameter in the Bose-Einstein condensed system (Fig.4). For the intermediate region, one analytically expects the spatial correlations to consist of a Gaussian thermal contribution $\exp(-4\pi|\mathbf{r}|^2/\lambda_{th}^2)$ at short length scales, as determined by the population in highly-excited transverse modes, and an exponentially decaying quasi long-range contribution $\exp(-2|\mathbf{r}|/\xi)$, which results from the macroscopic population of the low-energy states at quantum degeneracy[2,27]. Here, $\xi$ denotes the correlation length. In the thermodynamic limit, the described bimodal decay of correlations is expected only in the uncondensed phase ($N<N_c$), while upon condensation directly true long-range order, with $\xi \to \infty$, emerges. At photon numbers of order of $10^5$ for the here studied two-dimensional photon gas, however, finite-size effects become important, leading to a softening of the phase transition, which is associated with a continuous gradual increase of the correlation length. Correspondingly, the bimodal behavior is visible in our experimental and numerical data also for $N>N_c$, as seen in Fig.5. The corresponding expected population of low-energy states is given in the spectral photon distributions (inset).

From an exponential fit to the experimental data in Fig.5 between 2 and 6μm radial distance (dashed lines), we obtain a correlation length $\xi=7.4(2)$μm for a total photon number of 95000 ($N/N_c$=1.01) and $\xi$=9.6(5)μm for 96000 photons ($N/N_c$≈1.02), respectively. For both of these measurements the correlation length is of order of the size of the ground mode given by the oscillator length $\sigma_0=\sqrt{\hbar/(m\Omega)}$=7.7μm. Measurements closer to the condensation threshold of $N_c$=94000 in the harmonically trapped case would require a photon number precision on the order of 0.1%. For $N \lesssim N_c$, we cannot extract accurate experimental values for $g^{(1)}(|\mathbf{r}|)$ due to limited signal-to-noise ratio in the here required pulsed pumping operational mode and correspondingly we only show results of our numerical calculations. Nevertheless, our analysis reveals the predicted emergence of (quasi) long-range order in the regime close to the condensation threshold[2,27].

To conclude, we have determined spatial coherence properties of a two-dimensional photon gas, both in the thermal and the Bose-Einstein condensed phase. The high spatial resolution of the interferometric setup allows us to directly image the coherence properties of the photon gas. In the uncondensed regime, our measurements reveal that the extent of the photon wave packets is determined by the thermal de Broglie wavelength. We find excellent agreement both in the absolute value as well as the temperature scaling of the de Broglie wavelength with the expectations. We observe that quantum statistical effects, as indicated by long-range order, emerge when the thermal de Broglie wave packets spatially overlap, a behavior so far only verified for atomic gases. For the condensed phase, we find that the coherence extends over the whole sample. Close to the phase transition, the quantum degenerate gas exhibits thermal and quasi-long-range contributions to the spatial correlations as expected theoretically. To further explore this regime, it would be beneficial to study the photon gas confined in a box potential at large phase-space densities, which would allow a more quantitative comparison with theory predictions due to the absence of Bose-Einstein condensation in the homogeneous two-dimensional system[17,27,45]

For the future, spatially resolved first-order coherence measurements are expected to reveal possible long-lived phase singularities from vortices in thermo-optically or Kerr nonlinearity induced photon superfluids[46,47]. Other than atomic condensates, optical quantum gases can be subject to grand canonical statistics[48,49], which is expected to give rise to unusual dynamics of the quantum fluid. Finally, our setup might be a tool to study critical scaling at the phase transition.

## Methods

**Dye-cavity setup and characterization**

The dye microcavity uses two spherically curved high-reflecting dielectric mirrors (1m radius), as typically used in cavity ring-down spectroscopy with reflectivities >99.997% in the relevant wavelength regime (530-585nm). The mirrors are separated by $D_0$=1.63μm, corresponding to 4 optical wavelengths at 583nm in the solvent (refractive index $n$=1.43), causing a large frequency gap between adjacent longitudinal optical modes that is comparable to the emission bandwidth of the dye molecules. Thus, the resonator is populated only with photons of a fixed longitudinal mode ($q$=8) making the photon gas two-dimensional. The cavity is filled with Rhodamine 6G dye solved in ethylene glycol (dye concentration $10^{-3}$mol/l), acting as a heat bath for the photon gas. Collisions of solvent and dye molecules ($10^{-14}$s timescale) here suppress coherent energy exchange between photons and dye molecules, so that the photon gas is operated in the weak coupling regime[37]. The dye microcavity is pumped with a laser beam of ~100μm diameter at 532nm, exploiting a minimum in the mirror reflectivity at 43° angle to the optical axis. In the uncondensed regime pumping is done continuously with a pump power of ~1mW, yielding $N$=18(2) intracavity photons. However, as the optical pump power required to reach the critical photon number for condensation favors excessive population of long lived triplet states and photo bleaching of the organic dye molecules, continuous pumping is rendered unfeasible in the condensed phase. Therefore, using two acousto-optic modulators, the beam is chopped into pulses of 600ns length with a 50Hz repetition rate. To allow for a heating of the dye microcavity above room temperature, two electrical power resistors are glued to the cavity mount, allowing for a 12W thermal output. Further, two thermo-resistive temperature detectors are attached to the cavity mirror substrates from different cavity sides, and the quoted temperatures of the dye microcavity apparatus denote the average reading of these detectors with the error bar (Fig.3) reflecting the difference between the two readings.

The imaged dispersion shown in Fig.1b corresponds to a momentum-resolved spectrum of the cavity emission. For this, only the central ~30μm (diameter) of the cavity emission is transmitted through an iris aperture in the image plane, which effectively reduces the relative intensity contribution of the higher energy modes. The light subsequently passes a narrow slit in the momentum plane, before it is sent onto an optical grating (2400 lines/mm) and imaged onto the camera.

**Interferometer setup**

To characterize the first-order coherence of the photon gas, the emission on one side of the microcavity is collimated with a long working distance objective (Mitutoyo M Plan Apo 10x) and sent onto a Michelson-type interferometer. After passing a polarizer, the beam here is split up equally into two partial beams with a non-polarizing beam splitter, which are reflected by a plane mirror and a hollow cat-eye retroreflector, respectively. The latter is mounted on a piezo-driven stepper motor translator allowing for a steady variation of the path length difference of the interfering paths. The beams are recombined in the beam splitter following the usual Michelson interferometer arrangement, and then imaged on an EMCCD camera. The hollow mirror cat-eye retroreflector inverts the two transversal spatial coordinates (transverse with respect to the optical axis), and is illuminated off-center to bypass imperfections at the internal contacting edges of the device. Correspondingly, a slight tilt of the plane mirror back reflector is needed to match the two beams in the image plane. To observe interference fringes we sample images for different delay times, typically scanning the cat-eye retroreflector over a longitudinal distance of roughly 90μm, corresponding to a variation of the total time delay of 180μm/$c_0$≈600fs. During such a scan 2500 images are acquired, corresponding to time steps in the path difference near 0.2fs.

The obtained fringe visibility recorded for each of the camera pixels allows to generate a two-dimensional map of the correlation function $g^{(1)}(\mathbf{r},-\mathbf{r};\tau)$, where $\tau=\Delta\ell/c_0$, denotes the time delay accumulated due to the path length difference $\Delta\ell$. The transverse coherence length of the photon gas (Figs.2b,c and Fig.4) is obtained by evaluating fringe data with a path length difference much below the longitudinal coherence length, yielding effectively $g^{(1)}(\mathbf{r},-\mathbf{r};\tau)\equiv g^{(1)}(\mathbf{r},-\mathbf{r})$. The reduced maximal visibility in Fig.2c is attributed to the influence of detector noise floor (see Supplementary note 6). As the transverse coherence length of the uncondensed photon gas is of the same order of magnitude as the imaging resolution of the objective used to collimate the microcavity emission, the point spread function of the whole imaging system was carefully determined in preliminary measurements. To simulate the emission of a point source, the <200nm diameter aperture of a SNOM fiber tip (LovaLite EM50 SMF28) was placed in the emission plane of the cavity, the latter replaced by a non-reflecting cavity dummy of same shape and size, using a drop of ethylene glycol for index matching purposes. The resulting image of the fiber emission (utilizing a dye laser at 583nm) is used to characterize the complete imaging system. Both the spatial decay of the experimentally observed visibility data versus transverse position for a thermal photon gas, see Fig.2c, as well as the imaging system point spread function can be well approximated by Gaussian curves, with the latter exhibiting a width of $\sigma_{PSF} \simeq 0.658(2) \mu m$. A deconvolution of the measured Gaussian correlation signal of width $\sigma$ with the point spread function can thus readily be performed, and the true correlation length is obtained by $\sqrt{\sigma^2 - \sigma_{PSF}^2}$. The finite spatial resolution of the imaging system also implies that the temporal interference signal shown in Fig.2a corresponds to data averaged over a spatial area corresponding to the imaging resolution noted above, which is larger than the ~200x200nm size of one camera pixel in the imaging plane.

**Measurement details**

The measured absolute value of the coherence time of $\tau_c \simeq 129(6)$fs at $T$=297K temperature corresponds to $\tau_c = \kappa \hbar/(k_B T)$ with $\kappa$=4.90(3), which is larger than the expected $\kappa=\sqrt{3}$ for the case of perfect imaging resolution. This results from the finite spatial resolution, which mixes spatial correlations at different positions to yield the observed temporal correlation data (see Supplementary note 5). The data shown in Figs.2 and 3 were recorded at typical photon numbers in the dye microcavity apparatus of $N$=18(2), which is more than three orders of magnitude below the critical photon number $N_c$=94000.

The experimental data shown in Fig.4 investigates the condensed phase regime of the photon gas with condensate fractions ranging from 1% up to 55%, corresponding to ratios $N/N_c$ of up to 2.2(3). At the used total photon numbers, we find no evidence for a spatial broadening of the condensate cloud, as attributed to the here compared to Ref. 12 used longer spacing between successive pump pulses, so that thermo-optic interaction effects are considered to be small. The quoted values for $N/N_c$ in Fig.4 are determined from a spectroscopic measurement of the cavity emission using a spectrometer in $4f$ configuration, equipped with a motorized slit for wavelength selection and a photomultiplier. To allow for a comparison of the numerically obtained visibilities with the results extracted from the experimental data, we have accounted for the noise characteristics of the used detector, with an intrinsic noise floor present in the taken images. This effect reduces the resolvable visibility in the outer parts of the wings of the Gaussian condensate mode as the mean noise floor here becomes comparable to the expected cavity emission intensity, which reduces the obtained value for the coherence length. For details and a more elaborated discussion of the performed numerical simulations on correlations in a thermalized two-dimensional photon gas confined in a harmonic trap see the Supplementary notes 1-4.

The absolute photon number at criticality is found by evaluating spatial images of the emission recorded with a calibrated EMCCD camera. By considering the cavity characteristics (mirror transmission, round trip time, energy cutoff and pulse length), the transmission of all optical elements in the beam path and the detector characteristics (quantum efficiency, gain, electrons per count and detected areal fraction of the emission) with respective uncertainties, we experimentally find the critical intracavity particle number $N_c=90(18)\times10^3$ to be in good agreement with the theoretically expected value. We determine the critical phase-space density $\tilde{n}_c\lambda_{th}^2=3.2(6)$ from the central density $\tilde{n}=N/[\pi(\Delta r)^2]$, see Refs.[27,44], with the thermal cloud radius $\Delta r=\sqrt{2k_BT/(m_{eff}\Omega^2)}$. Note that the expected critical phase-space density $\tilde{n}_c\lambda_{th}^2=\pi^2/3\approx3.3$ differs from the value $\zeta(2)=\pi^2/6$ of the Riemann zeta function $\zeta$ for a two-dimensional system in a harmonic potential due to the intrinsic two-fold polarization degeneracy of the photonic system.

**References:**

1. Huang, K. *Statistical Mechanics* 2nd edn (Wiley, 1987).
2. Landau, L. D. & Lifshitz, E. M. *Statistical Physics* 3rd edn, Part 1: Vol. 5 (Butterworth-Heinemann, 1980).
3. Pethick, C. J. & Smith, H. *Bose–Einstein Condensation in Dilute Gases* 2nd edn (Cambridge University Press, 2008).
4. Naraschewski, M. & Glauber, R. J. Spatial coherence and density correlations of trapped Bose gases. *Phys. Rev. A* **59,** 4595–4607 (1999).
5. Pitaevskii, L. P. & Stringari, S. *Bose-Einstein Condensation* (Oxford University Press, 2003).
6. Anderson, M. H., Ensher, J. R., Matthews, M. R., Wieman, C. E. & Cornell, E. A. Observation of Bose-Einstein Condensation in a Dilute Atomic Vapor. *Science* **269,** 198–201 (1995).
7. Davis, K. B. *et al.* Bose-Einstein Condensation in a Gas of Sodium Atoms. *Phys. Rev. Lett.* **75,** 3969–3973 (1995).
8. Bradley, C. C., Sackett, C. A. & Hulet, R. G. Bose-Einstein Condensation of Lithium: Observation of Limited Condensate Number. *Phys. Rev. Lett.* **78,** 985–989 (1997).
9. Kasprzak, J. *et al.* Bose–Einstein condensation of exciton polaritons. *Nature* **443,** 409–414 (2006).
10. Balili, R., Hartwell, V., Snoke, D., Pfeiffer, L. & West, K. Bose-Einstein Condensation of Microcavity Polaritons in a Trap. *Science* **316,** 1007–1010 (2007).
11. Deng, H., Haug, H. & Yamamoto, Y. Exciton-polariton Bose-Einstein condensation. *Rev. Mod. Phys.* **82,** 1489–1537 (2010).
12. Klaers, J., Schmitt, J., Vewinger, F. & Weitz, M. Bose-Einstein condensation of photons in an optical microcavity. *Nature* **468,** 545–548 (2010).
13. Marelic, J. & Nyman, R. A. Experimental evidence for inhomogeneous pumping and energy-dependent effects in photon Bose-Einstein condensation. *Phys. Rev. A* **91,** 33813 (2015).
14. Andrews, M. R. *et al.* Observation of Interference Between Two Bose Condensates. *Science* **275,** 637–641 (1997).

15. Bloch, I., Hänsch, T. W. & Esslinger, T. Measurement of the spatial coherence of a trapped Bose gas at the phase transition. *Nature* **403,** 166–170 (2000).

16. Miller, D. E. *et al.* High-contrast interference in a thermal cloud of atoms. *Phys. Rev. A* **71,** 43615 (2005).

17. Donner, T. *et al.* Critical Behavior of a Trapped Interacting Bose Gas. *Science* **315,** 1556–1558 (2007).

18. Guarrera, V. *et al.* Observation of Local Temporal Correlations in Trapped Quantum Gases. *Phys. Rev. Lett.* **107,** 160403 (2011).

19. Dall, R. G. *et al.* Ideal n-body correlations with massive particles. *Nat. Phys.* **9,** 341–344 (2013).

20. Kavokin, A., Baumberg, J. J., Malpuech, G. & Laussy, F. P. *Microcavities*. (Oxford University Press, 2007).

21. Deng, H., Solomon, G. S., Hey, R., Ploog, K. H. & Yamamoto, Y. Spatial Coherence of a Polariton Condensate. *Phys. Rev. Lett.* **99,** 126403 (2007).

22. Wertz, E. *et al.* Spontaneous formation and optical manipulation of extended polariton condensates. *Nat. Phys.* **6,** 860–864 (2010).

23. Schmitt, J. *et al.* Spontaneous Symmetry Breaking and Phase Coherence of a Photon Bose-Einstein Condensate Coupled to a Reservoir. *Phys. Rev. Lett.* **116,** 33604 (2016).

24. Marelic, J. *et al.* Spatiotemporal coherence of non-equilibrium multimode photon condensates. *New J. Phys.* **18,** 103012 (2016).

25. Berezinskii, V. L. Destruction of Long-range Order in One-dimensional and Two-dimensional Systems Possessing a Continuous Symmetry Group. II. Quantum Systems. *J. Exp. Theor. Phys.* **34,** 610 (1972).

26. Kosterlitz, J. M. & Thouless, D. J. Ordering, metastability and phase transitions in two-dimensional systems. *J. Phys. C Solid State Phys.* **6,** 1181 (1973).

27. Hadzibabic, Z. & Dalibard, J. Two-dimensional Bose fluids: An atomic physics perspective. *Riv. Nuovo Cimento* **34,** 389 (2011).

28. Bagnato, V. & Kleppner, D. Bose-Einstein condensation in low-dimensional traps. *Phys. Rev. A* **44,** 7439–7441 (1991).

29. Masut, R. & Mullin, W. J. Spatial Bose-Einstein condensation. *Am. J. Phys.* **47,** 493–497 (1979).

30. Hadzibabic, Z., Krüger, P., Cheneau, M., Battelier, B. & Dalibard, J. Berezinskii–Kosterlitz–Thouless crossover in a trapped atomic gas. *Nature* **441,** 1118–1121 (2006).

31. Roumpos, G. *et al.* Power-law decay of the spatial correlation function in exciton-polariton condensates. *Proc. Natl. Acad. Sci.* **109,** 6467–6472 (2012).

32. Nitsche, W. H. *et al.* Algebraic order and the Berezinskii-Kosterlitz-Thouless transition in an exciton-polariton gas. *Phys. Rev. B* **90,** 205430 (2014).

33. Altman, E., Sieberer, L. M., Chen, L., Diehl, S. & Toner, J. Two-Dimensional Superfluidity of Exciton Polaritons Requires Strong Anisotropy. *Phys. Rev. X* **5,** 11017 (2015).

34. Dagvadorj, G. *et al.* Nonequilibrium Phase Transition in a Two-Dimensional Driven Open Quantum System. *Phys. Rev. X* **5,** 41028 (2015).

35. Klaers, J., Vewinger, F. & Weitz, M. Thermalization of a two-dimensional photonic gas in a 'white wall' photon box. *Nat. Phys.* **6,** 512–515 (2010).

36. Marelic, J., Walker, B. T. & Nyman, R. A. Phase-space views into dye-microcavity thermalized and condensed photons. *Phys. Rev. A* **94,** 63812 (2016).

37. De Martini, F. & Jacobovitz, G. R. Anomalous spontaneous stimulated-decay phase transition and zero-threshold laser action in a microscopic cavity. *Phys. Rev. Lett.* **60,** 1711–1714 (1988).

38. Kirton, P. & Keeling, J. Nonequilibrium Model of Photon Condensation. *Phys. Rev. Lett.* **111,** 100404 (2013).

39. Damm, T. *et al.* Calorimetry of a Bose–Einstein-condensed photon gas. *Nat. Commun.* **7,** 11340 (2016).

40. Schmitt, J. *et al.* Thermalization kinetics of light: From laser dynamics to equilibrium condensation of photons. *Phys. Rev. A* **92,** 11602 (2015).

41. Keeling, J. & Kirton, P. Spatial dynamics, thermalization, and gain clamping in a photon condensate. *Phys. Rev. A* **93,** 13829 (2016).

42. Loudon, R. *The Quantum Theory Of Light*. (Oxford University Press, U.S.A., 1997).

43. Kohnen, M. & Nyman, R. A. Temporal and spatiotemporal correlation functions for trapped Bose gases. *Phys. Rev. A* **91,** 33612 (2015).

44. Petrov, D. S., Gangardt, D. M. & Shlyapnikov, G. V. Low-dimensional trapped

gases. *J. Phys. IV Proc.* **116,** 5–44 (2004).

45. Gaunt, A. L., Schmidutz, T. F., Gotlibovych, I., Smith, R. P. & Hadzibabic, Z. Bose-Einstein Condensation of Atoms in a Uniform Potential. *Phys. Rev. Lett.* **110,** 200406 (2013).

46. de Leeuw, A.-W., Stoof, H. T. C. & Duine, R. A. Phase fluctuations and first-order correlation functions of dissipative Bose-Einstein condensates. *Phys. Rev. A* **89,** 53627 (2014).

47. Calvanese Strinati, M. & Conti, C. Bose-Einstein condensation of photons with nonlocal nonlinearity in a dye-doped graded-index microcavity. *Phys. Rev. A* **90,** 43853 (2014).

48. Klaers, J., Schmitt, J., Damm, T., Vewinger, F. & Weitz, M. Statistical Physics of Bose-Einstein-Condensed Light in a Dye Microcavity. *Phys. Rev. Lett.* **108,** 160403 (2012).

49. Schmitt, J. *et al.* Observation of Grand-Canonical Number Statistics in a Photon Bose-Einstein Condensate. *Phys. Rev. Lett.* **112,** 30401 (2014).

**Acknowledgements**

This work has been supported by the DFG (CRC 185) and the ERC (Inpec).

**Corresponding authors**

Correspondence to: Martin Weitz (martin.weitz@uni-bonn.de) or Julian Schmitt (schmitt@iap.uni-bonn.de).

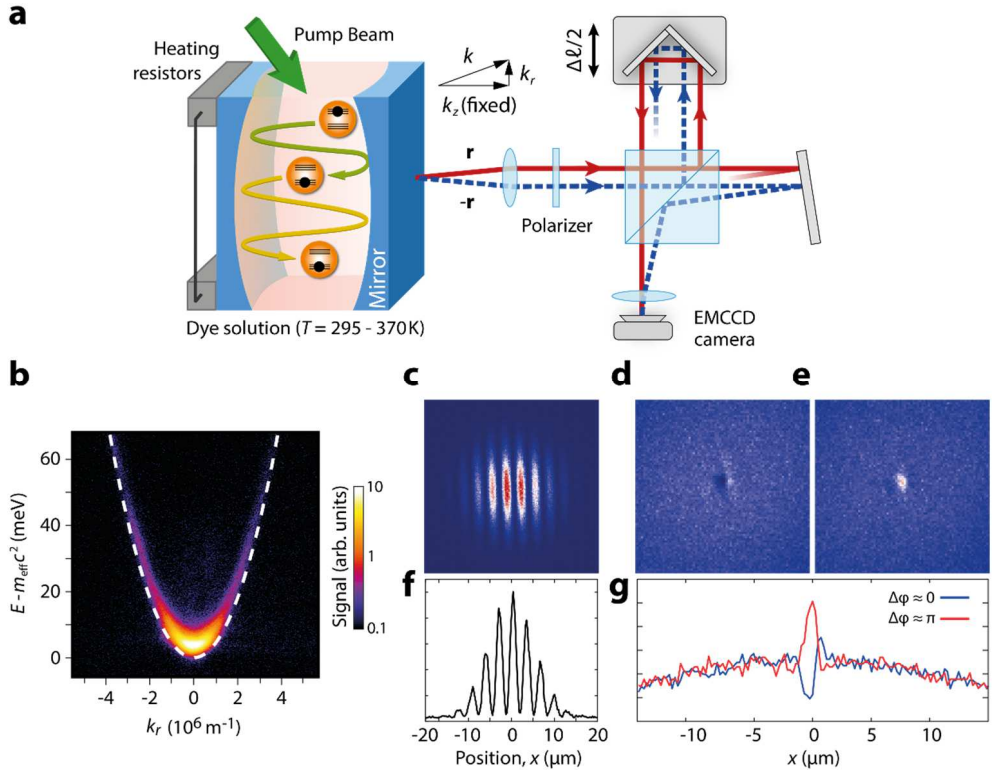

**Figure 1 | Experimental scheme. a.** The dye-filled optical microcavity confines photons between two highly-reflecting mirrors spaced in the wavelength regime, where they thermalize to the temperature of the dye solution by absorption re-emission processes. The ambient temperature of the apparatus can be varied between 297K (room temperature) and 370K. The Michelson interferometer shown on the right is used to investigate the transverse and longitudinal coherence of the photon gas. The sketched axial mirroring of transverse coordinates illustrates the actual point mirroring in the image plane done by the cat-eye retroreflector. **b.** Measured dispersion of the dye microcavity emission, showing photon energies versus the transverse wave vector $k_r$ (derived from the angle of emission). The data was recorded for the uncondensed gas. **c-g.** Camera images of the radiation transmitted by the interferometer in the condensed (**c**), and in the uncondensed (**d**, **e**) phase regime, with $N/N_c \approx 1.76$ and $N/N_c \approx 10^{-4}$, respectively. In the uncondensed phase, the transverse coherence length of the microcavity emission is so short that it amounts to less than a fringe width. The two camera images correspond to two different relative path lengths in the Michelson interferometer. The image size corresponds to 40x40μm (**c**) and 20x20μm (**d**, **e**), respectively. The bottom plots (**f**) and (**g**) give cuts through the center of the corresponding fringe patterns (**c**) and (**d**, **e**), respectively.

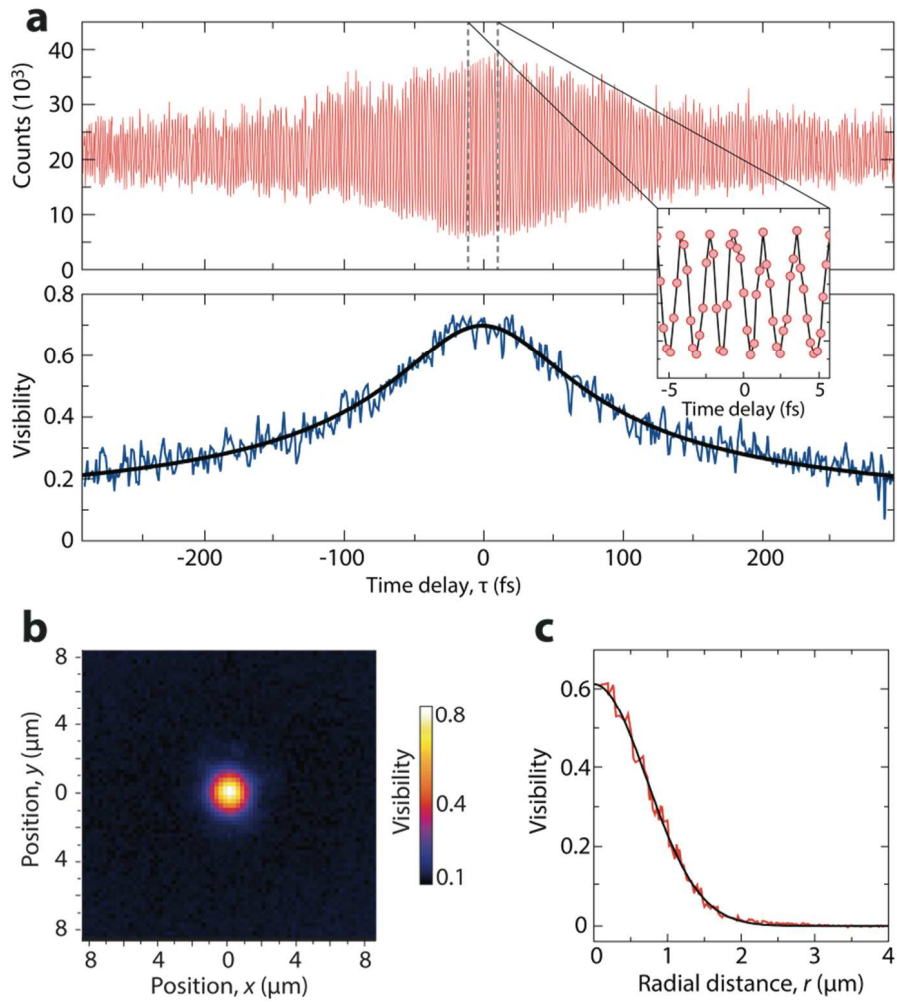

**Figure 2 | Temporal and spatial correlations of thermal photon gas. a.** The top panel shows the variation of the camera signal detected by the camera pixel closest to **r**=0 in the image plane as a function of the time delay due to different interferometer path lengths for a thermal photon gas ($N \ll N_c$). The inset gives a zoomed view into the central fringes. The bottom plot shows the corresponding variation of the fringe visibility along with a fit to $|g^{(1)}(\tau)|$ (solid). **b**. Map of the observed fringe visibility (raw data) versus transverse position in the camera plane. **c**. Offset corrected fringe visibility versus radial distance from the center (red), along with a Gaussian fit (black). After correcting for the measured resolution of the imaging system, the curve directly maps the extent of the thermal de Broglie wave packets. (T=307K, $N/N_c \simeq 10^{-4}$)

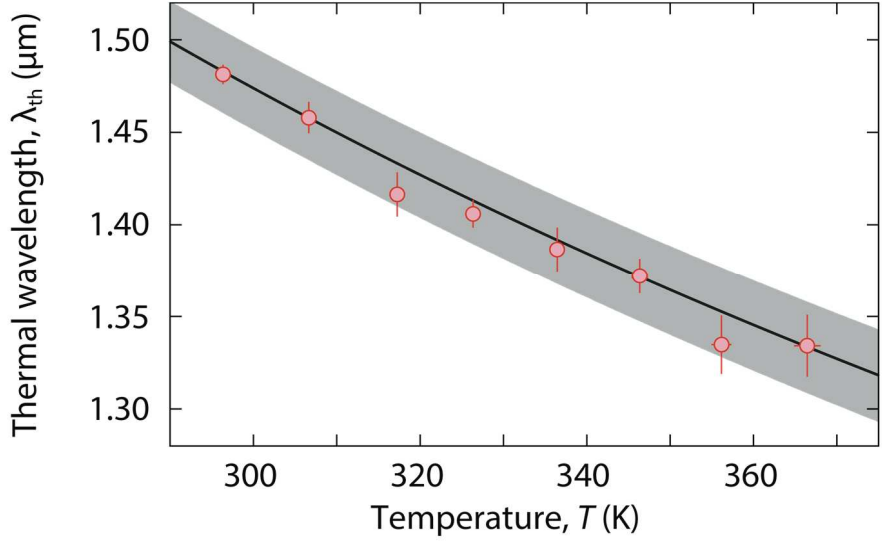

**Figure 3 | Temperature dependence of the thermal de Broglie wavelength.** Variation of the determined thermal de Broglie wavelength of the two-dimensional photon gas with the temperature of the dye microcavity apparatus. The data is in good agreement with the theory curve $\lambda_{th}=h/\sqrt{2\pi m_{eff}k_B T}$ (solid line), expected for a particle with mass $m_{eff}=\hbar\omega_{cutoff}/c^2$ at temperature $T$. The shaded area indicates the systematic uncertainty due to the correction for the measured finite imaging system resolution. The statistical uncertainty (s.d.) is indicated by the error bars.

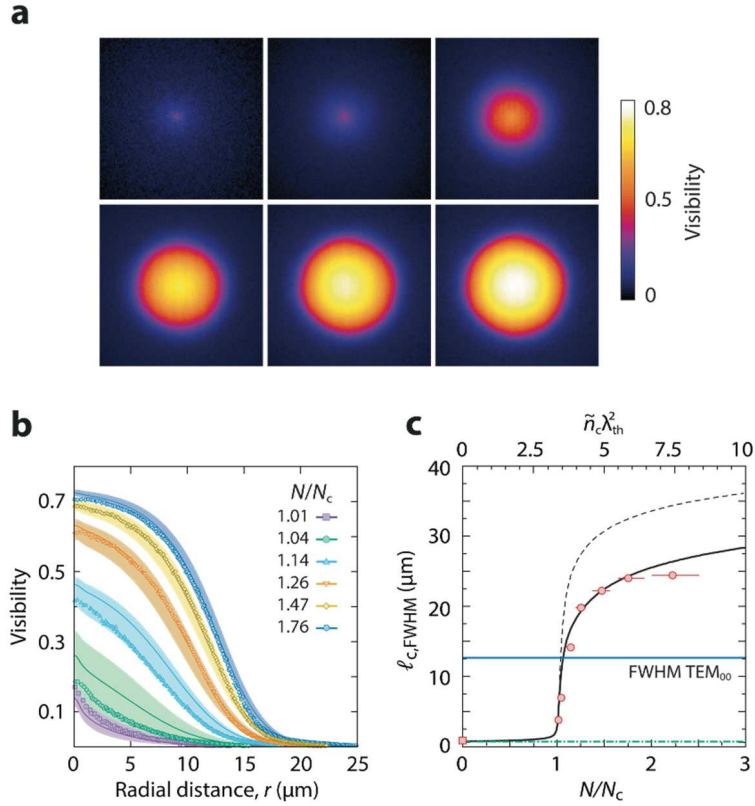

**Figure 4 | Emergence of coherence for a Bose-Einstein condensed photon gas. a.** Maps of the fringe visibility with transverse position for increasing ratios $N/N_c$ = {1.01; 1.04; 1.14; 1.26; 1.47; 1.76} showing a strong increase in the coherence, which indicates the onset of long-range order. The size of each image corresponds to 40x40μm. **b**. Corresponding variation of the fringe visibility (radially averaged) with distance from the center for different total photon numbers. The measured data (symbols) is accompanied by numerical simulations (solid lines) after taking into account the noise characteristics of the used EMCCD detector and the uncertainty in the experimental determination of the total photon number (shaded areas). **c**. Measured transverse coherence length, with the dots (square) corresponding to data recorded with pulsed (cw) pumping, respectively, versus $N/N_c$ (bottom axis) and phase-space density $\tilde{n}\lambda_{th}^2$ (top axis) along with theory for the coherence length with (dashed line) and without (solid) numerically implementing technical limitations. While much below the threshold to a Bose-Einstein condensate the transverse coherence length equals $\lambda_{th}\sqrt{\ln 2 /4\pi}$ (dash-dotted line), upon reaching Bose-Einstein condensation the ensemble becomes transversally coherent, with correlation lengths exceeding the 12.7μm FWHM of the condensate mode (blue solid line). The error bars represent the uncertainty (s.d.) in the determination of $N/N_c$.

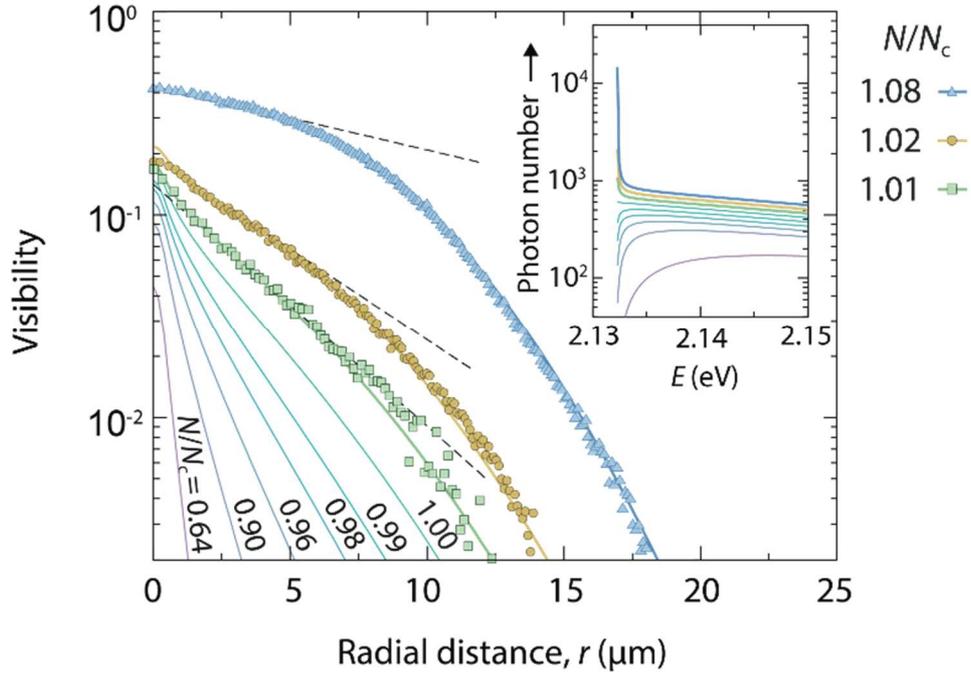

**Figure 5 | Correlations near the condensation threshold.** First-order correlations (symbols, see also Fig.4b) close to the critical photon number of $N_c$=94000 for three values of $N/N_c$, which have been determined by comparison with numerical calculations (solid lines). All three data sets follow closely the expected theoretical behavior. For $N/N_c$=1.01 (squares) and 1.02 (circles), we observe a short-range contribution to the correlations as attributed to Gaussian thermal correlations, while between 2 and 6μm the correlations reveal an exponential decay resulting from quantum degeneracy of low-energy states (see inset). For radii beyond the oscillator length $\sigma_0$=7.7μm, correlations decay due to the finite condensate width, as visible in the data for $N/N_c$=1.08 (triangles). Exponential fits (dashed lines) to the data yield correlation lengths $\xi$={7.4(2); 9.6(5); 21.6(29)}μm for $N/N_c$={1.01; 1.02; 1.08}. For the uncondensed phase ($N/N_c$<1), where data acquisition is prohibited by insufficient SNR of the imaging system, we show numerical calculations to illustrate the behavior of the photon gas correlations (solid lines). Inset: Theory spectra of the photon distribution (vertically shifted) for corresponding values of $N/N_c$ show the balanced population of the lowest-energy modes near the condensation point.

## Supplementary note 1: Introduction

In this Supplementary information, we give a theoretical study of the spatial and temporal first-order correlations of a two-dimensional photon gas confined in a harmonic trapping potential, assuming thermal equilibrium conditions. First, a rigorous analytical calculation for the first-order correlations of a classical (uncondensed) photon gas is presented. To test for the analytical model with its prediction for the transverse coherence length and the longitudinal coherence time, also numerical calculations are conducted. The numerical methods furthermore allow to extend predictions of the coherence properties to the Bose-Einstein condensed phase, where an analytical evaluation of the first-order correlation function fails. We investigate the scaling of the spatial and temporal first-order correlation length as a function of the total particle number and temperature, respectively. In the classical regime below the onset of Bose-Einstein condensation, the degree of first-order coherence is Gaussian, showing short-range correlations which decay on a length scale determined by the thermal de Broglie wavelength $\lambda_{\text{th}}$, see e.g. refs. 1,2. Our numerical results further verify the expected temperature dependence of $\lambda_{\text{th}} \propto 1/\sqrt{T}$ for the transverse coherence length and $\tau_{\text{c}} \propto 1/T$ for the longitudinal coherence time. In the quantum degenerate regime of the photon gas in the microcavity, when the lowest energy state is macroscopically occupied, our numerical results show the emergence of long-range correlations. The experimental observations presented in the main text are well described by the here discussed theory model.

## Supplementary note 2: Theory of spatial correlations

In this section, we calculate the first-order spatial correlations of a thermalized two-dimensional photon gas at temperature $T$ in a harmonic potential of angular trapping frequency $\Omega$, for total particle numbers well below the critical particle number for Bose-Einstein condensation $N_{\text{c}} = \pi^2/3 \left(k_{\text{B}} T/\hbar \Omega\right)^2$. To describe the experiments, we here theoretically study the first-order correlations $G^{(1)}(\mathbf{r}, \mathbf{r}')$ between two transverse positions $\mathbf{r}$ and $\mathbf{r}'$ in the photon gas at the same point in time. Accordingly, we will omit the explicit time dependence of the photon fields for the calculation of the transverse first-order correlations.

In the harmonic trapping potential the field operators $\hat{\Psi}$ and $\hat{\Psi}^\dagger$ can be expanded in harmonic oscillator eigenfunctions $\psi_n(\mathbf{r}) = \psi_n^*(\mathbf{r})$ and operators $\hat{a}_n, \hat{a}_n^\dagger$ for creation and annihilation, respectively, of a photon in an eigenstate with index $n$

$$\hat{\Psi}(\mathbf{r}) = \sum_m \psi_m(\mathbf{r}) \hat{a}_m, \quad \text{and} \quad \hat{\Psi}^\dagger(\mathbf{r}) = \sum_n \psi_n^*(\mathbf{r}) \hat{a}_n^\dagger. \tag{1}$$

The general expression for the first-order correlation function can be obtained from a Fourier transformation of the Wigner function, providing a connection between quantum-mechanical correlation function and statistical mechanics, see e.g. ref [1]. This yields

$$G^{(1)}(\mathbf{r}, \mathbf{r}') = \sum_n \psi_n^*(\mathbf{r}) \psi_n(\mathbf{r}') \langle \hat{a}_n^\dagger \hat{a}_n \rangle = \sum_n \psi_n^*(\mathbf{r}) \psi_n(\mathbf{r}') \bar{n}_n, \tag{2}$$

where we have used the orthogonality condition $\langle \hat{a}_m^\dagger \hat{a}_n \rangle = \delta_{mn}$ and the Bose-Einstein distributed average occupation of the $n$-th mode with energy $\epsilon_n$ at a chemical potential $\mu$ (regarding the two-fold polarization degeneracy per mode)

$$\langle \hat{a}_n^\dagger \hat{a}_n \rangle = \bar{n}_n = \frac{2}{e^{(\epsilon_n - \mu)/k_B T} - 1}. \tag{3}$$

In the dye microcavity the photon dynamics is restricted to the transverse degrees of freedom and discretized by the transverse modal quantum numbers $n_x$ and $n_y$. The energy levels of the two-dimensional harmonic oscillator are given by $E_{n_x n_y} = \hbar \omega_c + \hbar \Omega (n_x + n_y + 1)$, where $\hbar \omega_c$ is the cavity cutoff energy. With respect to this low-energy cutoff, the transverse energies in eq. (3) can be renormalized

$$\epsilon_{n_x n_y} := E_{n_x n_y} - \hbar \omega_c - \hbar \Omega = \hbar \Omega (n_x + n_y). \tag{4}$$

The corresponding spatially inhomogeneous eigenfunctions of the two-dimensional harmonic oscillator are given by

$$\psi_{n_x n_y}(x, y) = \frac{1}{\sqrt{2^{n_x} n_x!}} \frac{1}{\sqrt{2^{n_y} n_y!}} \left(\frac{m\Omega}{\pi\hbar}\right)^{1/2} e^{-\frac{m\Omega}{2\hbar}(x^2 + y^2)} H_{n_x}\left(\sqrt{\frac{m\Omega}{\hbar}} x\right) H_{n_y}\left(\sqrt{\frac{m\Omega}{\hbar}} y\right), \tag{5}$$

with $m$ as the photon mass and Hermite polynomials $H_n(x)$. The first-order correlation function then becomes

$$G^{(1)}(x, y, x', y') = \sum_{n_x, n_y} \frac{1}{2^{n_x} n_x!} \frac{1}{2^{n_y} n_y!} \frac{m\Omega}{\pi\hbar} e^{-\frac{m\Omega}{2\hbar}(x^2 + x'^2 + y^2 + y'^2)} \frac{2}{e^{(\epsilon_{n_x n_y} - \mu)/k_B T} - 1}$$
$$\times H_{n_x}\left(\sqrt{\frac{m\Omega}{\hbar}} x\right) H_{n_x}\left(\sqrt{\frac{m\Omega}{\hbar}} x'\right) H_{n_y}\left(\sqrt{\frac{m\Omega}{\hbar}} y\right) H_{n_y}\left(\sqrt{\frac{m\Omega}{\hbar}} y'\right) \tag{6}$$

In the thermal phase, where the chemical potential is well below thermal energy, $\mu \ll -k_B T$, the photon number distribution follows a Boltzmann distribution $\bar{n}_{n_x, n_y} \simeq 2 e^{\frac{\mu}{k_B T}} e^{-\frac{\hbar \Omega}{k_B T}(n_x + n_y)}$. In this case the correlation function at two identical positions $\mathbf{r} = (x, y)$ and $\mathbf{r}' = (x, y)$ simplifies to

$$G^{(1)}(x, y, x, y) = \frac{2m\Omega}{\pi\hbar} e^{\frac{\mu}{k_B T}} \times \left[ e^{-\frac{m\Omega}{\hbar} x^2} \sum_{n_x} \frac{e^{-\frac{\hbar\Omega}{k_B T} n_x}}{2^{n_x} n_x!} H_{n_x}\left(\sqrt{\frac{m\Omega}{\hbar}} x\right)^2 \right]$$
$$\times \left[ e^{-\frac{m\Omega}{\hbar} y^2} \sum_{n_y} \frac{e^{-\frac{\hbar\Omega}{k_B T} n_y}}{2^{n_y} n_y!} H_{n_y}\left(\sqrt{\frac{m\Omega}{\hbar}} y\right)^2 \right]. \tag{7}$$

Using the parameters $A = \frac{2m\Omega}{\pi\hbar} e^{\frac{\mu}{k_B T}}$, $\xi = \sqrt{\frac{m\Omega}{\hbar}} x$, $\eta = \sqrt{\frac{m\Omega}{\hbar}} y$, $\zeta = e^{-\frac{\hbar\Omega}{k_B T}}$ we rewrite eq. (7) as

$$G^{(1)}(\xi, \eta, \xi, \eta) = A \left[ e^{-\xi^2} \sum_{n_x} \frac{\zeta^{n_x}}{2^{n_x} n_x!} H_{n_x}(\xi)^2 \right] \times \left[ e^{-\eta^2} \sum_{n_y} \frac{\zeta^{n_y}}{2^{n_y} n_y!} H_{n_y}(\eta)^2 \right]. \tag{8}$$

In analogy, the correlation function for remote positions $r = (x, y)$ and $r' = (x', y')$ follows:

$$G^{(1)}(\xi, \eta, \xi', \eta') = A \left[ e^{-\frac{\xi^2+\xi'^2}{2}} \sum_{n_x} \frac{\zeta^{n_x}}{2^{n_x} n_x!} H_{n_x}(\xi) H_{n_x}(\xi') \right] \times \left[ e^{-\frac{\eta^2+\eta'^2}{2}} \sum_{n_y} \frac{\zeta^{n_y}}{2^{n_y} n_y!} H_{n_y}(\eta) H_{n_y}(\eta') \right]. \quad (9)$$

By employing the relation from ref. 3,

$$e^{-(\xi^2+\eta^2)} \sum_n \frac{\zeta^n}{2^n n!} H_n(\xi) H_n(\eta) = \frac{1}{\sqrt{1-\zeta^2}} e^{-\frac{\xi^2+\eta^2-2\xi\eta\zeta}{1-\zeta^2}}, \quad (10)$$

we from eqns. (8) and (9) obtain

$$G^{(1)}(\xi, \eta, \xi, \eta) = A \times e^{\xi^2+\eta^2} \frac{1}{1-\zeta^2} e^{-\frac{2(\xi^2+\eta^2)}{1+\zeta}} \quad (11)$$

$$G^{(1)}(\xi, \eta, \xi', \eta') = A \times e^{\frac{\xi^2+\xi'^2+\eta^2+\eta'^2}{2}} \frac{1}{1-\zeta^2} e^{-\frac{\xi^2+\xi'^2-2\xi\xi'\zeta}{1-\zeta^2}} \times e^{-\frac{\eta^2+\eta'^2-2\eta\eta'\zeta}{1-\zeta^2}}. \quad (12)$$

The normalized degree of first-order spatial coherence immediately follows to be

$$g^{(1)}(\xi, \eta, \xi', \eta') = \frac{G^{(1)}(\xi, \eta; \xi', \eta')}{\sqrt{G^{(1)}(\xi, \eta; \xi, \eta) \times G^{(1)}(\xi', \eta'; \xi', \eta')}} \quad (13)$$

$$= \exp\left[ -\frac{\zeta}{1-\zeta^2} \left( (\xi-\xi')^2 + (\eta-\eta')^2 \right) \right]. \quad (14)$$

By resubstituting the above defined parameters, one obtains the degree of first-order coherence of a thermalized classical two-dimensional photon gas

$$g^{(1)}(x, y, x', y') = \exp\left[ -\pi \frac{(\mathbf{r}-\mathbf{r}')^2}{\lambda_{\text{th}}^2} \right], \quad (15)$$

with the thermal de Broglie wavelength $\lambda_{\text{th}} = \hbar\sqrt{2\pi/mk_B T}$. This analytical result is exactly reproduced by our numerical calculations, see Supplementary Figure 1, which fulfill the same conditions as the experiment: the photon emission transmitted through one of the microcavity mirrors is split up and recombined with a retro-reflected copy of itself with inverted coordinates with respect to the origin, $(x, y) \leftrightarrow (-x, -y)$. With $\mathbf{r}' = -\mathbf{r}$ we obtain the degree of first-order coherence

$$g^{(1)}(r) = \exp\left[ -\frac{4\pi r^2}{\lambda_{\text{th}}^2} \right]. \quad (16)$$

Our numerical calculations (Supplementary Figure 1) are based on a Bose-Einstein distributed spectrum of photon energies and thus can be performed for different values of the chemical potential. In particular, this allows for a full numerical evaluation of the first-order correlations in the condensed phase for a large range of condensate fractions between 0% and close to 100%. The emergence of long-range order around the BEC threshold is visible in the two-dimensional maps showing $g^{(1)}(x, y, -x, -y)$ in Supplementary Figure 1(a), while Supplementary Figure 1(b) gives four corresponding spatial intensity distributions. Horizontal cuts through the correlation and intensity distributions at $y = 0$ are shown in Supplementary Figure 2(a). Below the critical particle number the spatial extent of the thermal correlations is in very good agreement with the analytical result in eq. (16), i.e. the correlations decay on a length scale determined by the thermal de Broglie wavelength of the photons at room temperature. On the other hand, in the Bose-Einstein-condensed phase our theory predicts an enhancement of the spatial first-order correlations to larger extensions, see also Supplementary Figure 2(c) for data of the transverse correlation length versus condensate fraction. In the thermal phase, the short-range correlation length exhibits a temperature dependence, as shown in Supplementary Figure 2(b), which can be understood from a temperature dependence of the size of the thermal de Broglie wavelength.

## Supplementary note 3: Theory of temporal correlations

In this section, the first-order temporal correlations of a trapped two-dimensional photon gas at thermal equilibrium conditions are calculated. As described above, the photons are confined in a harmonic trapping potential and the frequency of the photons in the transversal cavity modes is given by $E_{00}/\hbar = \omega_c + \Omega(n_x + n_y + 1)$. For the sake of simplicity, we consider correlations of fixed positions $(x, y) = (x', y')$ at different times $t, t'$, i.e. at non-vanishing time delays $\tau = t' - t \neq 0$,

$$G^{(1)}(x,y;t,t') = \sum_{n_x,n_y} \psi^*_{n_x n_y}(x,y) e^{-i[\omega_c+\Omega(n_x+n_y+1)]t} \psi_{n_x,n_y}(x,y) e^{i[\omega_c+\Omega(n_x+n_y+1)](n_x+n_y)t'} \bar{n}_{n_x,n_y}. \quad (17)$$

Far below the critical particle number, the first-order correlations of the thermal photon gas are

$$G^{(1)}(x,y;t,t+\tau) = \frac{2m\Omega}{\pi\hbar} e^{\frac{\mu}{k_B T}} \sum_{n_x,n_y} \frac{1}{2^{n_x} n_x! \, 2^{n_y} n_y!} H_{n_x}\left(\sqrt{\frac{m\Omega}{\hbar}}x\right)^2 H_{n_y}\left(\sqrt{\frac{m\Omega}{\hbar}}y\right)^2 \quad (18)$$

$$\times e^{-\frac{m\Omega}{\hbar}(x^2+y^2)} e^{i\Omega(n_x+n_y)\tau - \frac{\hbar\Omega}{k_B T}(n_x+n_y)} e^{i(\omega_c+\Omega)\tau} \quad (19)$$

$$= A e^{i(\omega_c+\Omega)\tau} \left[ e^{-\xi^2} \sum_{n_x} \frac{\zeta^{n_x}}{2^{n_x} n_x!} H_{n_x}(\xi)^2 \right] \times \left[ e^{-\eta^2} \sum_{n_y} \frac{\zeta^{n_y}}{2^{n_y} n_y!} H_{n_y}(\eta)^2 \right] \quad (20)$$

Similar to Section 1 of this Supplementary Information, we have used $A := \frac{2m\Omega}{\pi\hbar} e^{\frac{\mu}{k_B T}}$, $\xi := \sqrt{\frac{m\Omega}{\hbar}}x$, $\eta := \sqrt{\frac{m\Omega}{\hbar}}y$ and a redefined parameter $\zeta(\tau) := e^{i\Omega\tau - \frac{\hbar\Omega}{k_B T}}$, for which we will in the following steps drop the explicit notation of its time dependence, i.e. $\zeta(\tau) \equiv \zeta$. Using eq. (10) then yields

$$G^{(1)}(x,y;t,t+\tau) = A\, e^{i(\omega_c+\Omega)\tau} e^{\xi^2+\eta^2} \frac{1}{1-\zeta^2} e^{-\frac{2(\xi^2+\eta^2)}{1+\zeta}} \quad (21)$$

To obtain the first-order correlation function at equal times, we set $\tilde{\zeta} := e^{-\frac{\hbar\Omega}{k_B T}}$, where the dependence on the time delay $\tau$ has dropped out, and find

$$G^{(1)}(x,y;t,t) = G^{(1)}(x,y;t',t') = A\, e^{\xi^2+\eta^2} \frac{1}{1-\tilde{\zeta}^2} e^{-\frac{2(\xi^2+\eta^2)}{1+\tilde{\zeta}}}. \quad (22)$$

This readily yields the degree of first-order temporal coherence at a fixed position

$$g^{(1)}(x,y;t,t') = \frac{G^{(1)}(x,y;t,t')}{\sqrt{G^{(1)}(x,y;t,t)G^{(1)}(x,y;t',t')}} \quad (23)$$

$$= e^{i(\omega_c+\Omega)\tau}\frac{1-\tilde{\zeta}^2}{1-\zeta^2}\exp\left[-2\left(\xi^2+\eta^2\right)\left(\frac{\tilde{\zeta}-\zeta}{(1+\zeta)(1+\tilde{\zeta})}\right)\right]. \quad (24)$$

By resubstituting the parameters and using $r^2 = x^2 + y^2$ we obtain

$$g^{(1)}(r;\tau) = e^{i(\omega_c+\Omega)\tau}\frac{1-e^{-2\frac{\hbar\Omega}{k_BT}}}{1-e^{-2\frac{\hbar\Omega}{k_BT}+2i\Omega\tau}}\exp\left[-\frac{2m\Omega}{\hbar}r^2\left(\frac{1-e^{i\Omega\tau}}{1+e^{\frac{\hbar\Omega}{k_BT}}+e^{i\Omega\tau}+e^{-\frac{\hbar\Omega}{k_BT}+i\Omega\tau}}\right)\right]. \quad (25)$$

In the following, we omit the spatial dependence and consider the center of the photon gas at $r = 0$ only. Neither our experiments nor the ideal Bose gas theory indicate non-trivial propagating correlations, as were observed e.g. in polariton systems (see ref. 4). The first term in eq. (25) then determines the complex valued degree of first-order temporal coherence

$$g^{(1)}(\tau) = \frac{\left(e^{2\frac{\hbar\Omega}{k_BT}}-1\right)\{\cos\left[(\omega_c+\Omega)\tau\right]+i\sin\left[(\omega_c+\Omega)\tau\right]\}}{e^{2\frac{\hbar\Omega}{k_BT}}-\cos(2\Omega\tau)-i\sin(2\Omega\tau)}. \quad (26)$$

In the experiment one measures the real part of eq. (26)

$$\text{Re}\left[g^{(1)}(\tau)\right] = \frac{\left(e^{2\frac{\hbar\Omega}{k_BT}}-1\right)\left\{e^{2\frac{\hbar\Omega}{k_BT}}\cos\left[(\omega_c+\Omega)\tau\right]-\cos\left[(\omega_c-\Omega)\tau\right]\right\}}{1+e^{4\frac{\hbar\Omega}{k_BT}}-2e^{2\frac{\hbar\Omega}{k_BT}}\cos(2\Omega\tau)}, \quad (27)$$

which exhibits an oscillatory behaviour due to constructive and destructive interference between the photons from both arms of the Michelson interferometer at the detector. The degree of first-order coherence $\text{Re}\{g^{(1)}(\tau)\}+1$ is plotted in Supplementary Figure 3(a) as a function of the time delay for 4 different temperatures $T = \{100, 300, 380, 1000\}$K. The absolute of eq. (26) corresponds to the visibility shown in Supplementary Figure 3(b)

$$|g^{(1)}(\tau)| = \frac{\left(1-e^{-2\frac{\hbar\Omega}{k_BT}}\right)}{\sqrt{\left[1-e^{-2\frac{\hbar\Omega}{k_BT}}\cos(2\Omega\tau)\right]^2+\left[e^{-2\frac{\hbar\Omega}{k_BT}}\sin(2\Omega\tau)\right]^2}}, \quad (28)$$

which determines the coherence time $\tau_c$ as the time when the visibility has decayed to 0.5. With $a := \cosh\left(\frac{\hbar\Omega}{k_BT}\right)$ and $b := \sinh\left(\frac{\hbar\Omega}{k_BT}\right)$ it can be expressed by

$$\tau_c = \frac{1}{\Omega}\arccos\left\{\frac{\sqrt{-3+10a^2-3a^4-20ab+12a^3b+10b^2-18a^2b^2+12ab^3-3b^4}}{2(a-b)}\right\}. \quad (29)$$

As the inverse thermal excitation number $\hbar\Omega/k_BT \approx 0.007$ is well below unity in the dye microcavity,

an expansion up to first order in $\hbar\Omega/k_\mathrm{B}T$ provides a solution for the first-order correlation time

$$\tau_c \simeq \sqrt{3}\frac{\hbar}{k_\mathrm{B}T}\,. \tag{30}$$

Hence, the first-order (longitudinal) correlation time of the thermal photon gas is expected to scale as $1/T$, whereas the (transverse) spatial correlation length is expected to scale as $1/\sqrt{T}$, as given in eq. (16).

In the previously discussed derivation of the first-order temporal correlations of the thermal photon gas, the classical Boltzmann approximation $\bar{n}_{n_x n_y} \propto \exp\left(-\epsilon_{n_x n_y}/k_\mathrm{B}T\right)$ was used to obtain an analytical solution for $g^{(1)}(\tau)$. At the onset of Bose-Einstein condensation however, when the transverse ground mode occupation becomes macroscopic with respect to the total photon number, the first-order temporal correlations are significantly enhanced and the correlation function cannot be evaluated analytically any more. In this regime, the numerical calculations can be performed without restrictions and provide the exact first-order temporal correlation function by retaining the full Bose-Einstein distribution $\bar{n}_{n_x n_y} \propto \left[\exp\left(-\epsilon_{n_x n_y}/k_\mathrm{B}T\right) - 1\right]^{-1}$. Supplementary Figure 4 shows the visibility $|g^{(1)}(\tau)|$ at the center of the photon gas for 7 different values of the chemical potential and the total photon number, respectively, when crossing over from the thermal to the condensed regime at a temperature of 300K. Far below criticality, see the two lowest curves in Supplementary Figure 4, the coherence time of $\tau_c = 44\,\mathrm{fs}$ shows excellent agreement with the analytical result from eq. (30). As the total photon number is increased the coherence time becomes longer and above the critical particle number exceeds the investigated time range of 650fs.

**Supplementary note 4: Interference experiment**

In our experiment, the spatial and temporal correlations of the photon gas are studied by directing the light that is transmitted through one of the cavity mirrors into a Michelson-type interferometer, which allows both for a variation of the transverse displacement and the temporal delay between the interfering paths. One arm of the interferometer is equipped with a plane mirror, while the other arm hosts a cat-eye retroreflector mounted onto a linear motorized translation stage, with the latter spatially inverting the position coordinates of the photons. Both beam paths are recombined on an EMCCD camera with a quantum efficiency of approximately 97%. The spatial resolution of the used optical system is roughly $0.93\,\mu\mathrm{m}$, as we have verified by placing a point-like optical source (SNOM fiber, 200 nm diameter) in the microcavity plane, where the thermal photon gas is generated. In this section, we will briefly summarize the expected spatio-temporal intensity profile at the position of the detector.

The quantized photon field operators in eq. (1) can be extended by temporally varying phase terms

$$\hat{\Psi}(\mathbf{r}',t') = \sum_m \psi_m(\mathbf{r}')e^{-\frac{i}{\hbar}E_m t'}\hat{a}_m, \quad \text{and} \quad \hat{\Psi}^\dagger(\mathbf{r},t) = \sum_n \psi_n^*(\mathbf{r})e^{\frac{i}{\hbar}E_n t}\hat{a}_n^\dagger. \tag{31}$$

Correspondingly, the first-order correlation function at a temporal delay $\tau = t' - t$ due to a difference in the interferometer arm lengths becomes

$$G^{(1)}(\mathbf{r}, \mathbf{r}'; \tau) = \langle \hat{\Psi}^\dagger(\mathbf{r}, t) \hat{\Psi}(\mathbf{r}', t') \rangle = \sum_k \psi_k^*(\mathbf{r}) \psi_k(\mathbf{r}') e^{-\frac{i}{\hbar} E_k \tau} \, \bar{n}_k. \tag{32}$$

As the spatial intensity distribution of the photon gas in the microcavity, see ref. 5, is given by

$$I(\mathbf{r}) \simeq \frac{\hbar \omega_c}{\tau_{rt}} \sum_k |\psi_k(\mathbf{r})|^2 \bar{n}_k = \frac{\hbar \omega_c}{\tau_{rt}} G^{(1)}(\mathbf{r}, \mathbf{r}; 0) = \frac{1}{2} \varepsilon_0 c \langle |E(\mathbf{r}, t)|^2 \rangle, \tag{33}$$

one can make the heuristic ansatz for the operator of the electric field

$$\hat{E}(\mathbf{r}, t) = \sqrt{\frac{2\hbar \omega_c}{\varepsilon_0 c \tau_{rt}}} \hat{\Psi}(\mathbf{r}, t) = \sqrt{\frac{2\hbar \omega_c}{\varepsilon_0 c \tau_{rt}}} \sum_m \psi_m(\mathbf{r}) e^{-\frac{i}{\hbar} E_m t} \hat{a}_m. \tag{34}$$

Assuming that the cavity emission of spatial intensity distribution $I(\mathbf{r})$ is incident on a symmetric beam splitter with $|r_i|^2 = |t_i|^2 = \frac{1}{2}$, the superposition of the electric fields in both the reflecting and spatially inverting interferometer arms results in an intensity at position $\mathbf{r}$ at the detector

$$\begin{aligned}
I_d(\mathbf{r}, \tau) &= \frac{1}{2} \varepsilon_0 c \langle | \frac{1}{2} E(\mathbf{r}, t) + \frac{1}{2} E(-\mathbf{r}, t') |^2 \rangle \\
&= \frac{1}{8} \varepsilon_0 c \frac{2\hbar \omega_c}{\varepsilon_0 c \tau_{rt}} \left\{ \sum |\psi_m(\mathbf{r})|^2 \bar{n}_m + \sum |\psi_m(-\mathbf{r})|^2 \bar{n}_m + 2 \operatorname{Re} \left[ \sum_m \psi_m^*(\mathbf{r}) \psi_m(-\mathbf{r}) \, e^{-\frac{i}{\hbar} E_m \tau} \bar{n}_m \right] \right\} \\
&= \frac{1}{4} \left\{ I(\mathbf{r}) + I(-\mathbf{r}) + 2\sqrt{I(\mathbf{r}) I(-\mathbf{r})} \operatorname{Re} \left[ g^{(1)}(\mathbf{r}, -\mathbf{r}; \tau) \right] \right\}
\end{aligned} \tag{35}$$

Here $\langle ... \rangle$ denotes a temporal averaging over times which are long compared to oscillation period of the optical field ($\tau = \lambda_c / c \simeq 2\,\text{fs}$, for $\lambda_c = 583\,\text{nm}$) caused by the detector and the degree of first-order coherence is given by

$$g^{(1)}(\mathbf{r}, -\mathbf{r}; \tau) = \frac{\langle \hat{E}^\dagger(\mathbf{r}, t) \hat{E}(-\mathbf{r}, t') \rangle}{\sqrt{\langle \hat{E}^\dagger(\mathbf{r}, t) \hat{E}(\mathbf{r}, t) \rangle} \sqrt{\langle \hat{E}^\dagger(-\mathbf{r}, t) \hat{E}(-\mathbf{r}, t) \rangle}} = \frac{\sum_m \psi_m^*(\mathbf{r}) \psi_m(-\mathbf{r}) \, e^{-\frac{i}{\hbar} E_m \tau} \bar{n}_m}{\sqrt{\sum_m |\psi_m(\mathbf{r})|^2 \bar{n}_m} \sqrt{\sum_m |\psi_m(-\mathbf{r})|^2 \bar{n}_m}}.$$

The real part of $g^{(1)}(\mathbf{r}, -\mathbf{r}; \tau)$ varies like an amplitude-modulated cosine function with an in general complicated phase $\Delta\phi(\omega_m, k_m)$, accounting for the thermal distribution of eigenfrequencies and a potential wavevector mismatch in both arms $(k_m - k_m')x$. Our final expression for the spatial intensity distribution at the output of the Michelson interferometer after rewriting eq. (35) is

$$I_d(\mathbf{r}, \tau) = \frac{1}{4} \left\{ I(\mathbf{r}) + I(-\mathbf{r}) + 2\sqrt{I(\mathbf{r}) I(-\mathbf{r})} \, |g^{(1)}(\mathbf{r}, -\mathbf{r}; \tau)| \, \cos\left[ \Delta\phi(\omega_m, k_m) \right] \right\}. \tag{36}$$

The absolute of the degree of first-order coherence thus determines the visibility of the interference fringes, which are measured in the experiment:

$$V(\mathbf{r}, \tau) = \frac{I_{\max} - I_{\min}}{I_{\max} + I_{\min}} = \frac{2\sqrt{I(\mathbf{r}) I(-\mathbf{r})}}{I(\mathbf{r}) + I(-\mathbf{r})} |g^{(1)}(\mathbf{r}, -\mathbf{r}; \tau)| \tag{37}$$

## Supplementary note 5: Effects of finite spatial resolution

The finite spatial resolution of the optical imaging system used in the experiment modifies the measured correlation signal, which has to be accounted for in the numerical calculations. Assuming a radially symmetric intensity distribution $I(\mathbf{r}) = I(-\mathbf{r})$, the real part and the absolute of the degree of first-order temporal coherence at any position $\mathbf{r}$ read

$$\mathrm{Re}\left[g^{(1)}(\mathbf{r},\tau)\right] = \frac{4I_d(\mathbf{r},\tau)}{2I(\mathbf{r})} - 1 \quad \text{and} \quad |g^{(1)}(\mathbf{r},\tau)| = V(\mathbf{r},\tau), \tag{38}$$

which give the degree of temporal coherence at an infinitesimal spatial position. Experimentally however one measures the intensity $\tilde{I}_d(\mathbf{r},\tau)$, which is averaged over the vicinity of the position under investigation. The width of the averaging is determined by the spatial resolution of the imaging apparatus; more precisely, it is quantified by the width of the point spread function on the detector produced by a point-like source placed in the resonator plane, where the thermal photon gas is generated. Numerically, this can be accounted for by integrating over the nearby detection area at each given time delay,

$$\begin{aligned}
\tilde{I}_d(\mathbf{r},\tau) &= \frac{1}{A}\int_{\mathrm{PSF}} I_d(\mathbf{r}',\tau)\exp\left[-\frac{(\mathbf{r}-\mathbf{r}')^2}{2\sigma_{\mathrm{PSF}}^2}\right]d\mathbf{r}' \\
&\simeq \frac{I(\mathbf{r},\tau)}{2}\left\{1 + \frac{1}{A}\int_{\mathrm{PSF}} |g^{(1)}(\mathbf{r}',-\mathbf{r}';\tau)|\exp\left[-\frac{(\mathbf{r}-\mathbf{r}')^2}{2\sigma_{\mathrm{PSF}}^2}\right]d\mathbf{r}'\right\},
\end{aligned}$$

with the normalization factor $A = \int_{\mathrm{PSF}} \exp\left[-\frac{(\mathbf{r}-\mathbf{r}')^2}{2\sigma_{\mathrm{PSF}}^2}\right]d\mathbf{r}'$. Due to its slow variation with position, $I(\mathbf{r},\tau)$ can be considered constant over a length scale given by the standard deviation $\sigma_{\mathrm{PSF}}$ of the point spread function, and thus it is excluded from the integral. Finally, one obtains the following expected experimentally detectable degree of first-order coherence

$$|\tilde{g}^{(1)}(\mathbf{r},-\mathbf{r};\tau)| = \frac{2\tilde{I}_d(\mathbf{r},\tau)}{I(\mathbf{r},\tau)} - 1. \tag{39}$$

The numerical result for the real part of the spatially averaged first-order correlation $\mathrm{Re}\left[g^{(1)}(0,0;\tau)\right]$ for an uncondensed, thermal photon gas with $\mu \simeq -7.0 k_\mathrm{B} T$ is shown in Supplementary Figure 5(a). The data demonstrates that the visibility of the averaged interference fringes (green line) is reduced at short time delays, as compared to the visibility of the interference fringes with perfect spatial resolution (blue line). At large time delays, the absolute of the averaged first-order correlation function approaches the non-averaged visibility, as remote points with $|\mathbf{r}| > 0$ here exhibit no significant contributions due to the finite spatial coherence of the photons. Similarly, when approaching the condensed phase regime (see Supplementary Figure 5(b) for three increasing values of the chemical potential), no significant difference between the averaged (solid line) and non-averaged (thick solid line) visibility is observed, as is well understood from the here long-range spatial correlations which by far exceed the imaging resolution. By integrating the numerically calculated interference signals over an area equivalent to that of the point spread function, the numerical results well reproduce our experimentally measured signals observations (grey lines in Supplementary Figure 5(a,b)). The finite imaging resolution has to be particularly considered if one measures the spatial correlations at zero path difference in the interferometer ($\tau \simeq 0$). Supplementary Figure 5(c) shows the analytical result for the first-order spatial correlations (thick blue line) from eq. (16), the numerical result with (dashed line) and without (red line) a convolution with the experimentally determined point spreading function with $\sigma_{\text{PSF}} = 0.658(2)\,\mu\text{m}$, along with the experimental data (yellow circles). The experimentally observed data are in excellent agreement with the theoretical averaged correlation function.

## Supplementary note 6: Effects of detector characteristics

Our experimental procedure to obtain two-dimensional maps of spatial correlations relies on a position-resolved measurement of the fringe visibility in an interferometer with nearly equal arm lengths. As visible e.g. in Figure 1(c) of the main text, the density of the harmonically trapped photon gas is not spatially homogeneous, but rather due to the effect of the trapping potential concentrated around the trap center. Correspondingly, while the visibility of the interference pattern can be determined reliably in this dense central region, a determination of the interference contrast in the low-intensity regions is much more imprecise. The finite noise floor of the used camera now in the outer spatial regions with a small signal turns out to be additive, as our camera always gives only positive signal outputs, while the effect of the noise is both positive and negative for a large camera signal, as obtained near the trap center. For illustration purposes, Supplementary Figure 6 shows a cut through the camera signal recorded in the absence of a fringe pattern, which shows the described effect of (asymmetric) symmetric noise in the (low intensity) high intensity region, respectively. For the data with a fringe pattern, the presence of the asymmetric noise floor of the camera $\delta I$ in the outer parts of the photon cloud with $I_{\text{max,min}} \to I_{\text{max,min}} + \delta I$ effectively reduces the visibility value $V_{\text{exp}}$ of the interference fringes derived from the observed camera signals,

$$V(\mathbf{r}) = \frac{I_{\text{max}} - I_{\text{min}}}{I_{\text{max}} + I_{\text{min}}} =: \frac{A}{B}, \quad V_{\text{exp}}(\mathbf{r}) = \frac{A}{B + 2\delta I}. \quad (40)$$

An expansion in $2\delta I$ yields

$$V_{\text{exp}}(\mathbf{r}) = V(\mathbf{r}) \left[ 1 - \frac{\delta I}{\bar{I}(\mathbf{r})} + \left(\frac{\delta I}{\bar{I}(\mathbf{r})}\right)^2 - \left(\frac{\delta I}{\bar{I}(\mathbf{r})}\right)^3 + ... \right] = V(\mathbf{r}) \left(1 + \frac{\delta I}{\bar{I}(\mathbf{r})}\right)^{-1}, \quad (41)$$

with $I_{\text{max}} + I_{\text{min}} \approx 2\bar{I}(\mathbf{r})$, where $\bar{I}(\mathbf{r})$ denotes the average intensity in the vicinity of position $\mathbf{r}$. While for large intensities the derived value for the visibility approaches the ideal value, for low intensity levels the obtained visibility value is significantly reduced. From a measurement of the spatial intensity distribution, we find significant deviations between experimental and theoretical data for photon densities near $3 \cdot 10^{13}\,\text{m}^{-2}$, see Supplementary Figure 6. The effect of this detector noise floor has been accounted for in the numerical model calculation of the expected fringe contrast. The corresponding expected curves can be found in Figures 4(b),(c) of the main text, which results in a much better agreement with the experimentally determined visibility values.

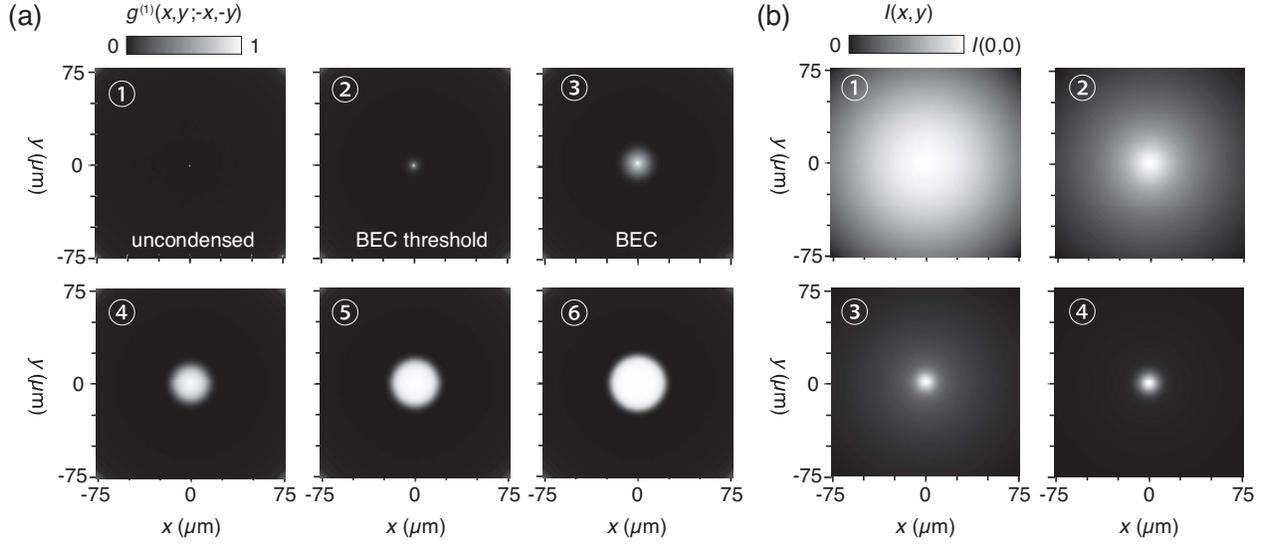

**Supplementary Figure 1**: **Numerical results for correlations and intensity distributions.** (a) First-order spatial correlations $g^{(1)}(x,y;-x,y)$ for increasing values of the chemical potential $\mu_{①\text{-}⑥} = -\{10^0; 10^{-2}; 10^{-3}; 10^{-4}; 10^{-5}; 10^{-6}\}\,k_B T$ (relative ground state occupation of $(n_0/N)_{①\text{-}⑥} = \{0.01\%; 0.2\%; 2\%; 19\%; 70\%; 96\%\}$) when crossing over from the thermal to the Bose-Einstein condensed phase. The chemical potential at criticality ($N_c = 94\,000$) is given by approximately $\mu_② \simeq -0.01\,k_B T$. In the thermal phase, correlations are short-range and determined by the thermal de Broglie wavelength. In the condensed phase, full spatial coherence is established within the condensate mode volume. (b) Spatial intensity distributions of the photon gas for corresponding values of $\mu_{①\text{-}④}$, both below and above critical photon number. All numerical calculations include contributions from the first 700 energy levels.

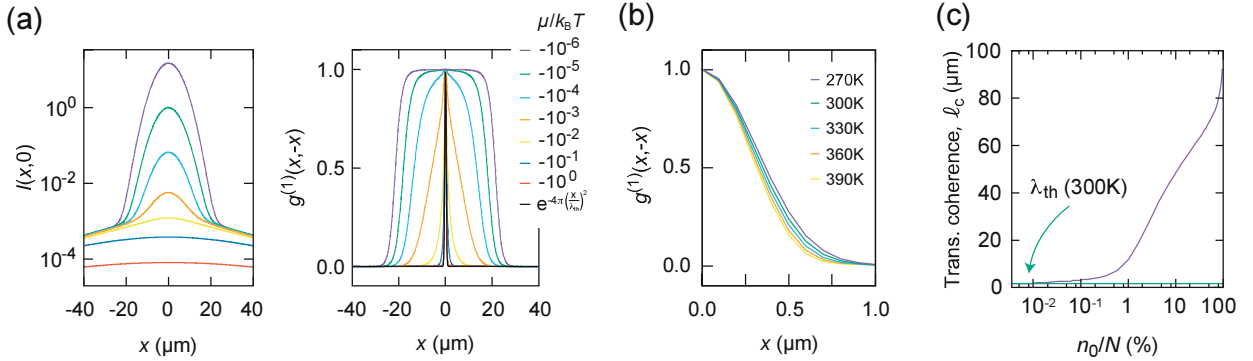

**Supplementary Figure 2**: **Spatial profiles of intensity distribution and first-order correlations.** (a) $I(x,0)$ and $g^{(1)}(x,-x)$ along $x$-axis for different values of the chemical potential ranging from the thermal to the Bose-Einstein-condensed phase. The solid black line corresponds to a theory curve based on eq. (16). (b) Transverse spatial correlations of the thermal photon gas (with $N = \text{const.}$) for different temperatures. When the photon gas is heated up, its correlation length decreases due to a reduction of the thermal de Broglie wavelength of the photons. (c) Extension of the transverse coherence length (FWHM) versus condensate fraction. Below the critical photon number, the coherence decays on a length scale characteristic for the thermal de Broglie wavelength $\lambda_{\text{th}}(300\text{K}) = 1.48\,\mu m$, while the correlation length increases when the threshold for Bose-Einstein condensation is exceeded.

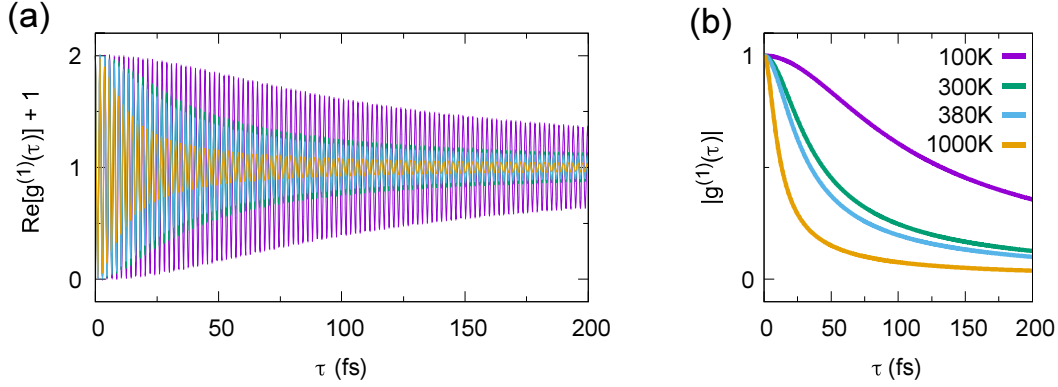

**Supplementary Figure 3**: **Temperature dependence of the degree of first-order temporal coherence.** (a) Real part of the degree of coherence as a function of time delay for 4 different temperatures based on the analytical derivation presented in this Supplementary Information. (b) Absolute of the degree of coherence, which corresponds to the visibility in an interference experiment with observed signal of type (a), for respective temperatures. For all plots, the photon gas is assumed to be in thermal equilibrium with a total particle number far below the critical number for Bose-Einstein condensation.

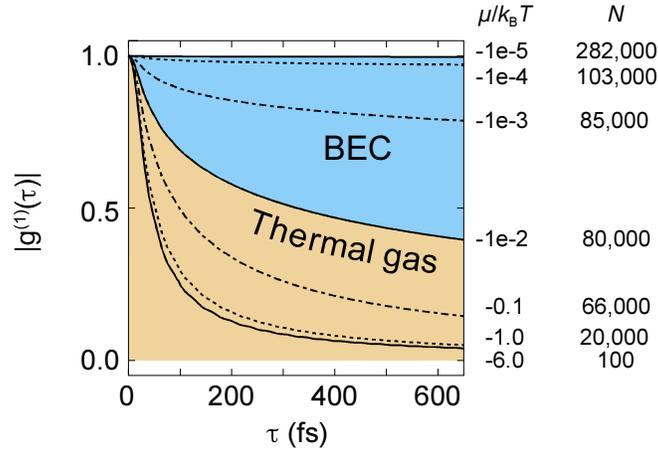

**Supplementary Figure 4**: **Numerical calculation of the degree of first-order temporal coherence for different photon numbers.** Far below condensation threshold (lowest two curves), the coherence decays on a time scale given by the analytical formula of eq. (30), which yields $\tau_c \simeq 44\,\text{fs}$ for 300K in perfect agreement with our numerics. With the onset of a Bose-Einstein condensate, long-range temporal coherence is established, which very soon above the threshold exceeds the shown time range of 650 fs. Note that above $\mu/k_\text{B}T \simeq -10^{-2}$ one crosses from the thermal to the Bose-Einstein condensed phase. Besides the chemical potential also the corresponding total number of particles $N$ is given.

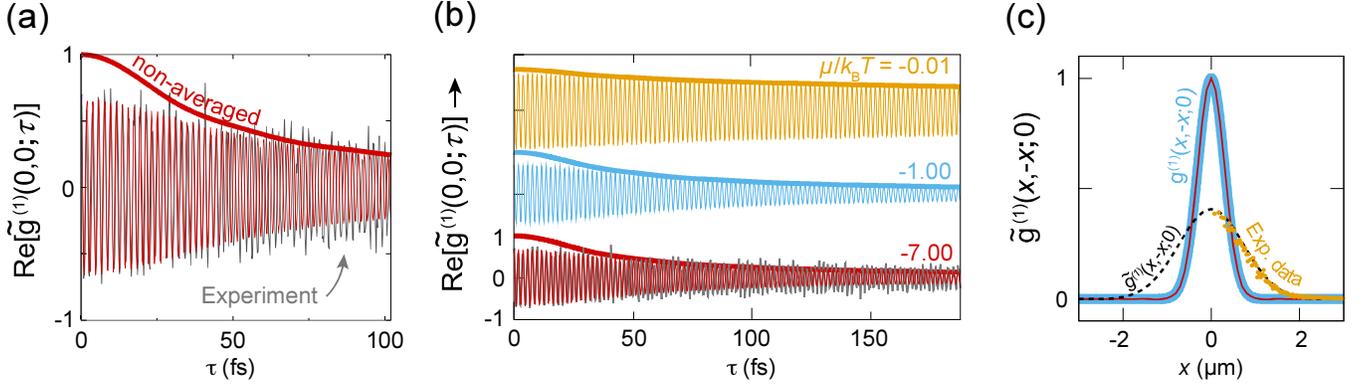

**Supplementary Figure 5**: **Effects of finite spatial resolution.** (a) Experimentally observed interference signal versus time delay (grey) along with numerically calculated interference signal (red) averaged over an area determined by the point spread function. The averaging leads to a reduction of the visibility of the interference fringes below the non-averaged visibility at the central position of the photon gas (thick red line). (b) Numerically calculated averaged interference signals versus time delay as the photon gas number is increased towards Bose-Einstein criticality. The effect of averaging is mostly visible at short time delays, and nearly vanishes as the chemical potential is increased towards criticality ($\mu/k_\mathrm{B}T \simeq -0.01$). (c) Spatial first-order correlations of a thermal photon gas at room temperature along the $x$-axis. The analytical result (blue) agrees with the numerical result (red). The finite imaging resolution in the experiment can be accounted for in the numerical calculations by convolving the spatial first-order correlation function with the point spread function (dashed line), demonstrating very good agreement between experiment and theory.

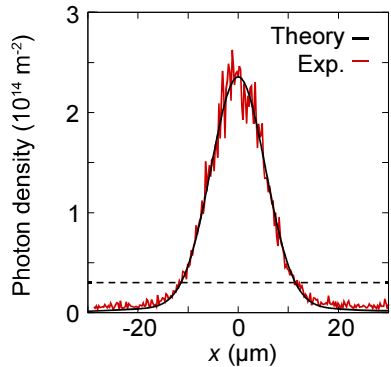

**Supplementary Figure 6**: **Detection noise floor.** Horizontal cut through the detected spatial intensity distribution from a single-shot measurement of a Bose-Einstein condensed photon gas in the absence of a fringe pattern (red line), along with theory for the photon density (black line). For photon densities below roughly $3 \cdot 10^{13}\,\mathrm{m}^{-2}$ (dashed line) the experimental data reveals a noise floor which exceeds the expected photon distribution. For the data with a fringe pattern, these effects cause a correction to the derived fringe visibility, in contrast to the observation in the trap center with large signal, where the noise is both positive and negative.

**Supplementary References**

[1] Naraschewski, M. & Glauber, R. J. Spatial coherence and density correlations of trapped Bose gases. *Phys. Rev. A* **59**, 4595–4607 (1999).

[2] Hadzibabic, Z. & Dalibard, J. Two-dimensional Bose fluids: An atomic physics perspective. *Riv. Nuovo Cimento* **34**, 389 (2011).

[3] Sakurai, J. J. *Modern Quantum Mechanics* revised edn (Addison-Wesley, 1994).

[4] Roumpos, G. et al. Power-law decay of the spatial correlation function in exciton-polariton condensates. *Proc. Natl. Acad. Sci.* **109**, 6467–6472 (2012).

[5] Klaers, J., Schmitt, J., Vewinger, F. & Weitz, M. Bose-Einstein condensation of photons in an optical microcavity. *Nature* **468**, 545–548 (2010).